\def\TITLE{Properties of a spectroscopically selected CV sample}

\documentstyle[11pt,newpasp,twoside,psfig]{article}
\begin{document}

\title{\TITLE}

\author{Boris T.~G\"ansicke}
\affil{Universit\"ats-Sternwarte,\,Geismarlandstr.\,11,\,37083\,G\"ottingen,\,Germany}

\author{H.-J.~Hagen, D.~Engels}
\affil{Hamburger Sternwarte, Gojenbergsweg 112, 21029 Hamburg, Germany}

\def\Porb{\mbox{$\mathrm{P_{orb}}$}}

\begin{abstract}
We have initiated a dedicated search for new CVs, selecting candidates
on the base of their spectroscopic properties in the Hamburg Quasar
Survey (HQS), which lead up to now to the identification of 50 new CVs
and a dozen good candidates. Using the HQS data of the previously
known CVs for extensive tests of our selection scheme, we demonstrate
that our survey should be very efficient in finding short period
systems as long as they have emission lines with equivalent
widths $\ga10$\,\AA. So far, orbital periods have been measured for 15
of the new CVs, with the surprising result that only two systems lie
at or below the lower boundary of the period gap.  This --~somewhat
preliminary~-- result is in uncomfortable disagreement with the
predictions of the standard scenario of CV evolution.
\end{abstract}

\section{The standard CV evolution scenario}
\begin{figure}[t]
\begin{minipage}{6.5cm}
\psfig{file=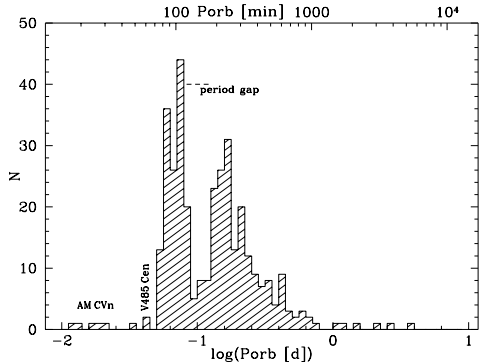,angle=-90,width=6.5cm}
\hspace*{-9mm}
\parbox{8cm}{\caption{Orbital period distribution of 347 CVs,
taken from TPP (Kube et al., this volume, p.\pageref{paper_kube_tpp}).
\label{f-cvporb}}}
\end{minipage}
\begin{minipage}{6.5cm}
\vspace*{-4ex}
Table\,1. CV space densities derived from the observations

\medskip
\begin{tabular}{rl}
\noalign{\smallskip}
\hline
\noalign{\smallskip}
$\ga10^{-6}\mathrm{pc^{-3}}$       & Warner 1974 \\
$5.7\times10^{-6}\mathrm{pc^{-3}}$ & Patterson 1984 \\
$8.2\times10^{-7}\mathrm{pc^{-3}}$ & Downes 1986\\
$2.1\times10^{-5}\mathrm{pc^{-3}}$ & Hertz et al. 1990\\
$6.0\times10^{-6}\mathrm{pc^{-3}}$ & Ringwald 1996\\
$1.4\times10^{-6}\mathrm{pc^{-3}}$ & Thomas \& \\
                                   & Beuermann 1998$^{*}$ \\
\noalign{\smallskip}
\hline
\hline
\noalign{\smallskip}
\end{tabular}
\smallskip
\hspace*{2cm}$^{*}$ {\small Only AM Herculis stars!}
\end{minipage}
\vspace*{-2ex}
\end{figure}

For almost two decades, the general approach to the theory of CV
evolution and the observational picture co-existed quite peacefully
despite some disagreements (Patterson 1984), and the progress on both
fronts was steady, but slow.
The best-known binary parameter of a CV is typically its orbital
period, and much of the theoretical effort has been devoted to a
quantitative description of the observed orbital period distribution
(Fig.\,\ref{f-cvporb}). The pronounced features of this distribution are (1) a
dwindling number of CVs at long orbital periods, (2) a dearth of systems
in the $2-3$\,h orbital period range, and (3) a sharp cut-off at
$\sim80$\,min. 
The standard CV evolution scenario (e.g., King 1988), which
qualitatively explains all three features, is based on the assumption
that two different angular momentum loss (AML) mechanisms are
dominating above (magnetic braking) and below (gravitational
radiation) the gap. While this standard model has recently been
challenged on various grounds (see in this volume:
Collier-Cameron, p.\,\pageref{paper_colliercameron}; Kolb,
p.\,\pageref{paper_kolb_bouncing}; King \& Schenker,
p.\,\pageref{paper_king}; Schenker \& King,
p.\,\pageref{paper_schenker}), we will focus here only on one aspect
in which the predictions of the theory disagree with the observations:
the space density of CVs.

Population syntheses based on the standard CV evolution scenario
predict rather large space densities of CVs, ranging from a few
$10^{-5}\,\mathrm{pc}^{-3}$ (de Kool 1992) to a few
$10^{-4}\,\mathrm{pc}^{-3}$ (Politano 1996)\footnote{To illustrate these
numbers: a space density of $10^{-5}\,\mathrm{pc}^{-3}$ corresponds to
42 CV within a distance of 100\,pc.}. Because the assumption of
enhanced AML above the gap implies a much shorter evolutionary time
scale for the long period CVs, the vast majority ($\ga95$\,\%) of the
present day CV population should be short period systems below the gap
(Kolb 1993; Howell, Rappaport, \& Politano 1997).

On the observational side, most space densities derived from different
CV samples are \textit{significantly} lower than the theoretical
values. Only the value of Hertz et al. (1990) is marginally in the
predicted range. However, their result is based on the identification of
four optically faint CVs in a rather small ($144\,\mathrm{deg}^2$)
area of the Einstein Galactic Plane Survey, and for none of these CVs
a reliable distance or orbital period measurement is available so far.

The number of Thomas \& Beuermann (1998) deserves an interesting note:
A space density of $1.4\times10^{-6}\,\mathrm{pc}^{-3}$ is derived for
the AM\,Herculis stars alone, CVs containing a strongly magnetic white
dwarf, on the base of the ROSAT All Sky Survey (RASS). If the ratio
magnetic/non-magnetic white dwarfs in CVs is the same as it is for
field white dwarfs ($2-5$\,\%, Jordan 1997; Wickramasinghe \&
Ferrario 2000), the result for the AM\,Her stars predicts a \textit{total}
CV space density of $\sim3-7\times10^{-5}\,\mathrm{pc}^{-3}$, which
implies \textit{independently of the standard scenario} that we are
missing a large fraction of the non-magnetic CVs.

While Fig.\,\ref{f-cvporb} gives only the period distribution of the
well-studied CVs which are, somewhat generalized, the brightest ones,
it suggests that we are predominantly missing short period CVs. An
exact assessment of the actual space densities above and below the gap
requires (a) good measurements of the distances, which are available
only for a limited number of systems, and (b) a detailed model for the
discovery probability. Especially the second point is a challenge
(e.g. Ritter 1986), as the selection effects depend on a large number
of parameters.

\section{How to find CVs}
CVs are discovered by various means. As suggested by their name,
\textit{variability} is one of their main features, this is especially true
for classical nova and dwarf novae with sudden brightness variations
of up to 10 magnitudes. Consequently, the bulk of the known CVs has
been discovered by legions of amateur observers relentlessly scanning
the skies for new stars, or by large-scale photographic sky patrols. 
Because the mass transfer from the secondary star onto the white dwarf
releases a large amount of potential energy, CVs can be detected as
X-ray emitters, in particular magnetic CVs, in which the bulk of the
energy release is confined to a rather small volume near the magnetic
pole caps of the white dwarf (polars, intermediate polars).  The
various X-ray satellite missions have discovered somewhat more than
100 CVs, the majority during the RASS (e.g. Thomas et al. 1998;
Beuermann et al. 1999).
As CVs typically contain a \textit{hot} component (accretion
disc/column, white dwarf), they show up as blue/ultraviolet excess
objects in colour surveys, such as the Palomar-Green (Green et
al. 1982; Ringwald 1996) survey. However, the colour space confusion
of CVs with other blue objects (white dwarfs, subdwarfs, QSOs) makes
the definite classification of CV candidates very telescope time
consuming.  To date, less than 100 CVs have been discovered in pure
colour surveys.
Finally, CVs may be discovered thanks to their particular
spectroscopic appearance: most CVs show relatively strong Balmer and
helium emission lines. However, so far this selection criterion has
been used only to a very limited extent, most of the 2--3 dozen
spectroscopically discovered CVs stem from the first and second
Byurakan surveys.

\section{The Hamburg Quasar Survey\label{s-hqs}}
The Hamburg Quasar Survey (HQS, Hagen et al. 1995) is a wide-angle
objective-prism survey to search for bright ($B \la17.5$) quasars in
the northern sky.  The survey was carried out from 1980 to 1997 with
the 80\,cm Schmidt telescope on Calar Alto, covering
$\approx13\,600\,\mathrm{deg}^2$ of the northern sky ($\delta>0\deg$)
at high galactic latitudes ($|b|>20\deg$).  The dynamic range of the
survey is $13 \la B \la 18.5$ and the spectral coverage is
$3400-5400$\,\AA\ with a resolution of $\sim45$\,\AA\ at $H_{\gamma}$.
The plates were scanned with a PDS 1010 G microdensitometer in
low-resolution mode with $\sim15$ pixel per spectrum. Blue objects
were selected among the low-resolution spectra and subsequently
rescanned with full resolution. The full-resolution density spectra
were classified after a visual inspection into a small number of
categories, primarily candidates for quasars, hot stars, and narrow
emission line objects (Hagen et al. 1999).  These spectra
($\sim50000$) form our archive of high-resolution scan (HRS) spectra
from which the candidates for our ongoing search for new CVs are
selected.


\begin{figure}[t]
\noindent
\mbox{\psfig{file=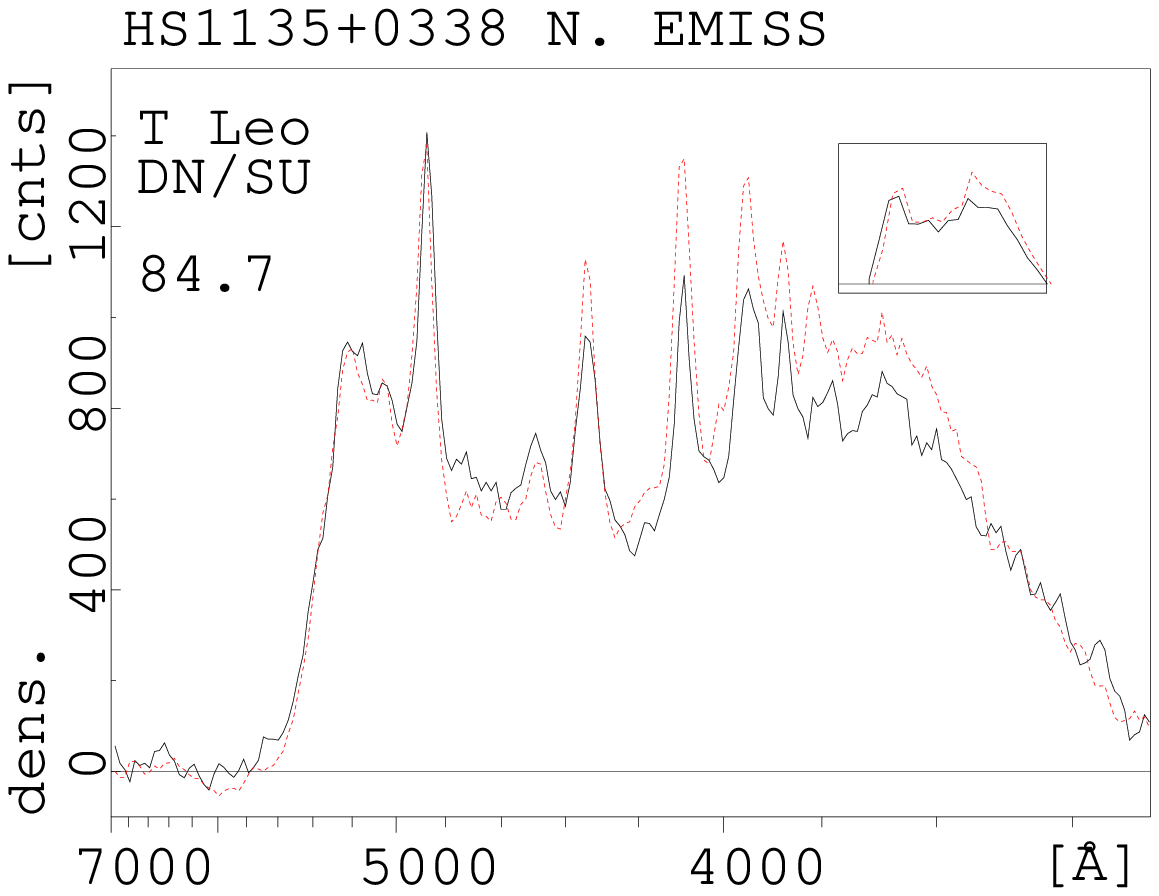,angle=-90,width=4.35cm}}
\mbox{\psfig{file=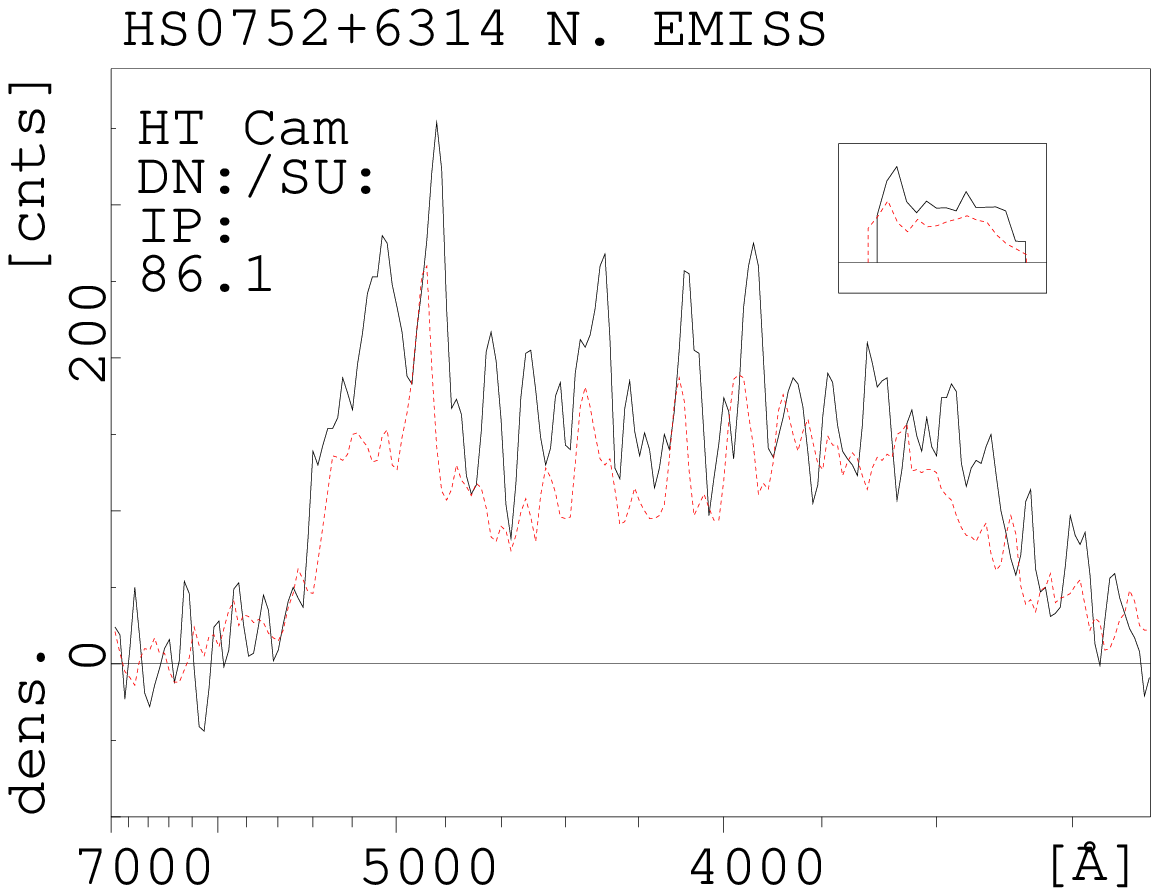,angle=-90,width=4.35cm}}
\mbox{\psfig{file=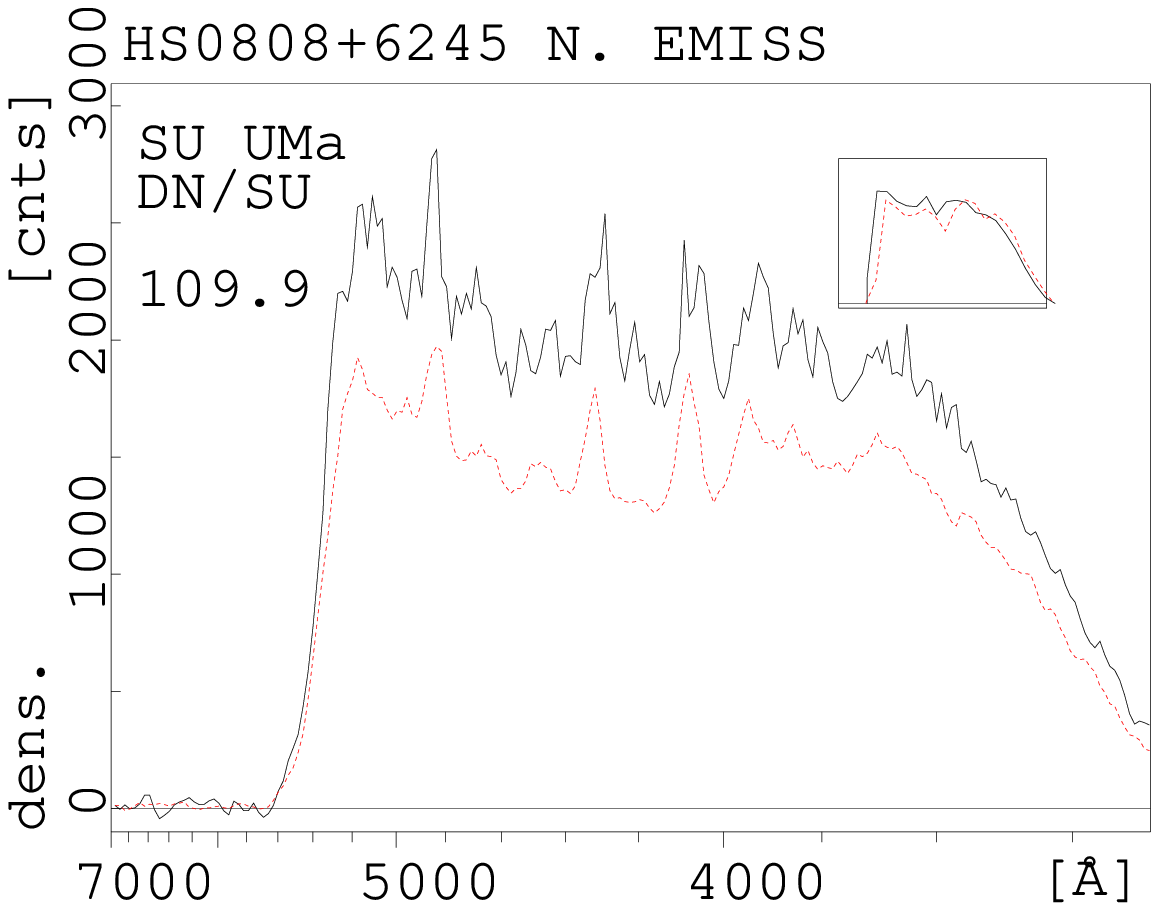,angle=-90,width=4.35cm}}

\mbox{\psfig{file=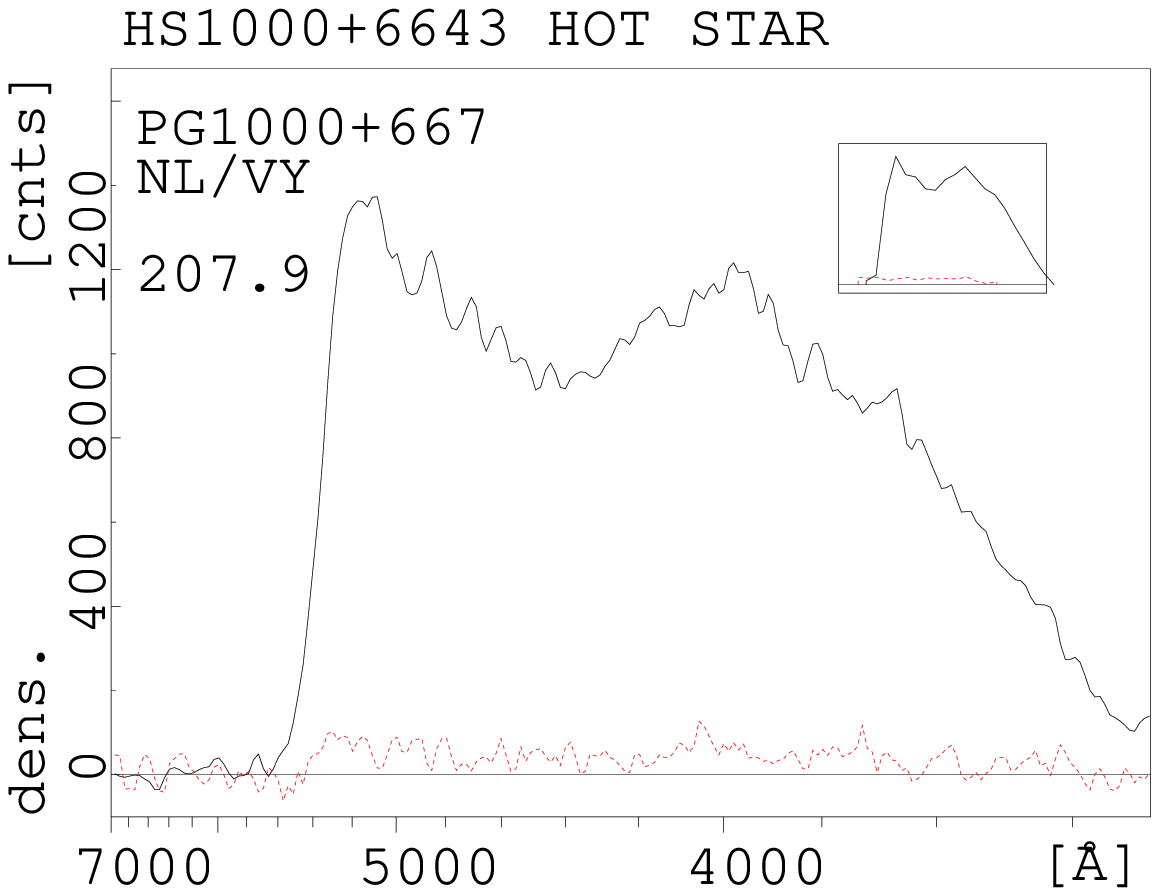,angle=-90,width=4.35cm}}
\mbox{\psfig{file=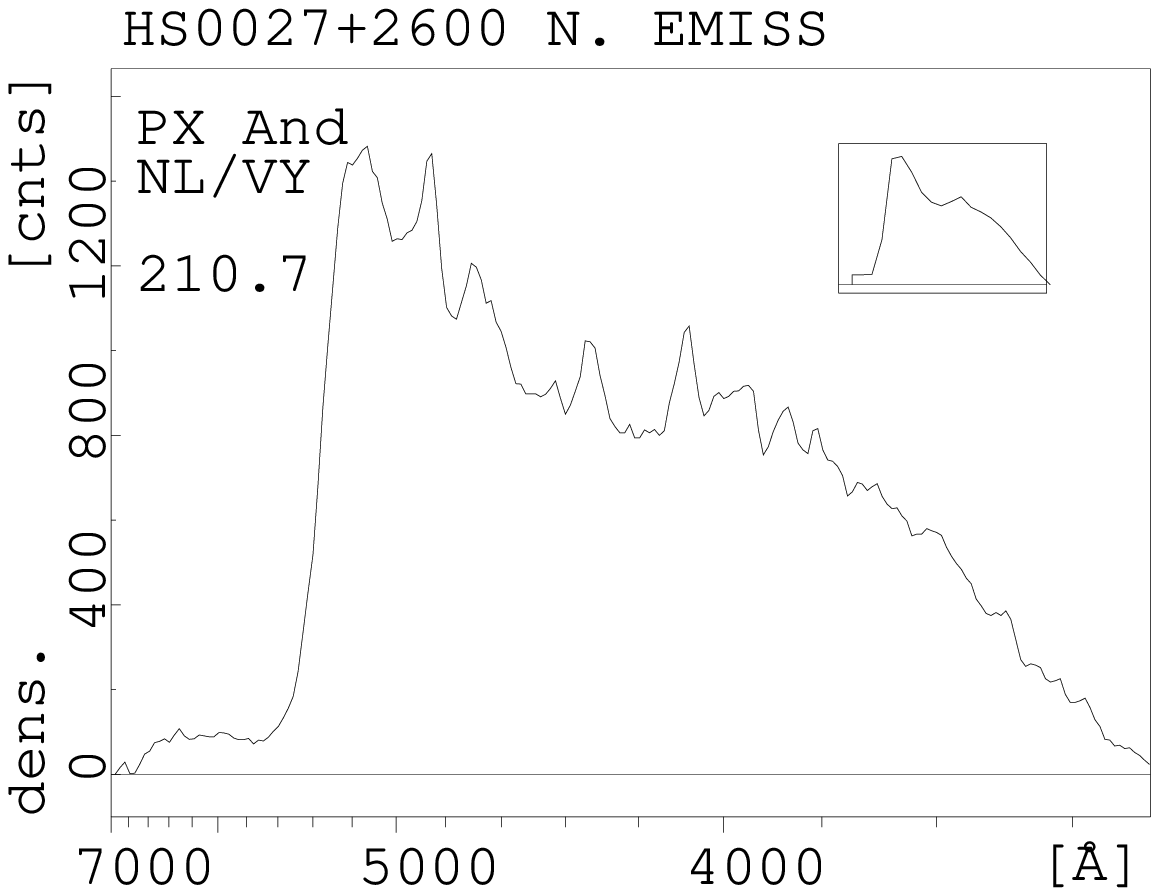,angle=-90,width=4.35cm}}
\mbox{\psfig{file=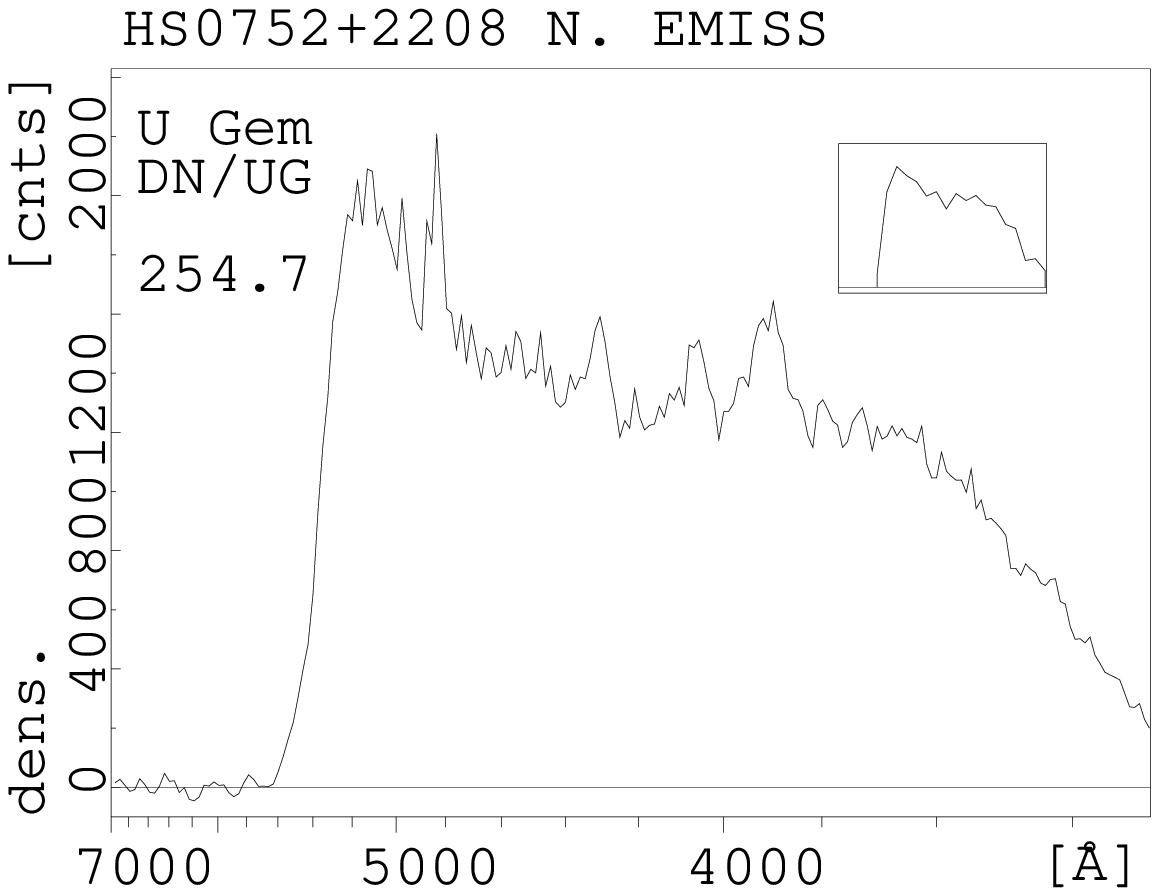,angle=-90,width=4.35cm}}

\mbox{\psfig{file=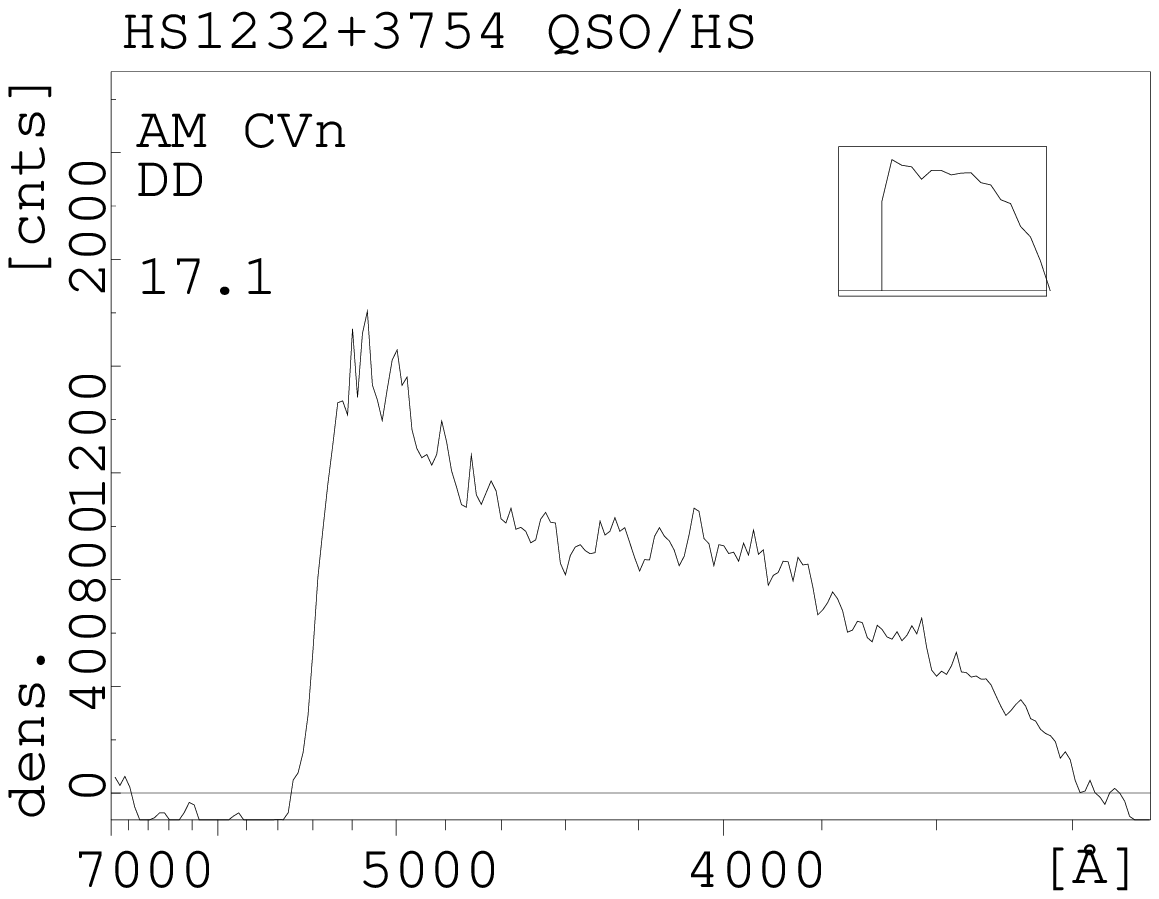,angle=-90,width=4.35cm}}
\mbox{\psfig{file=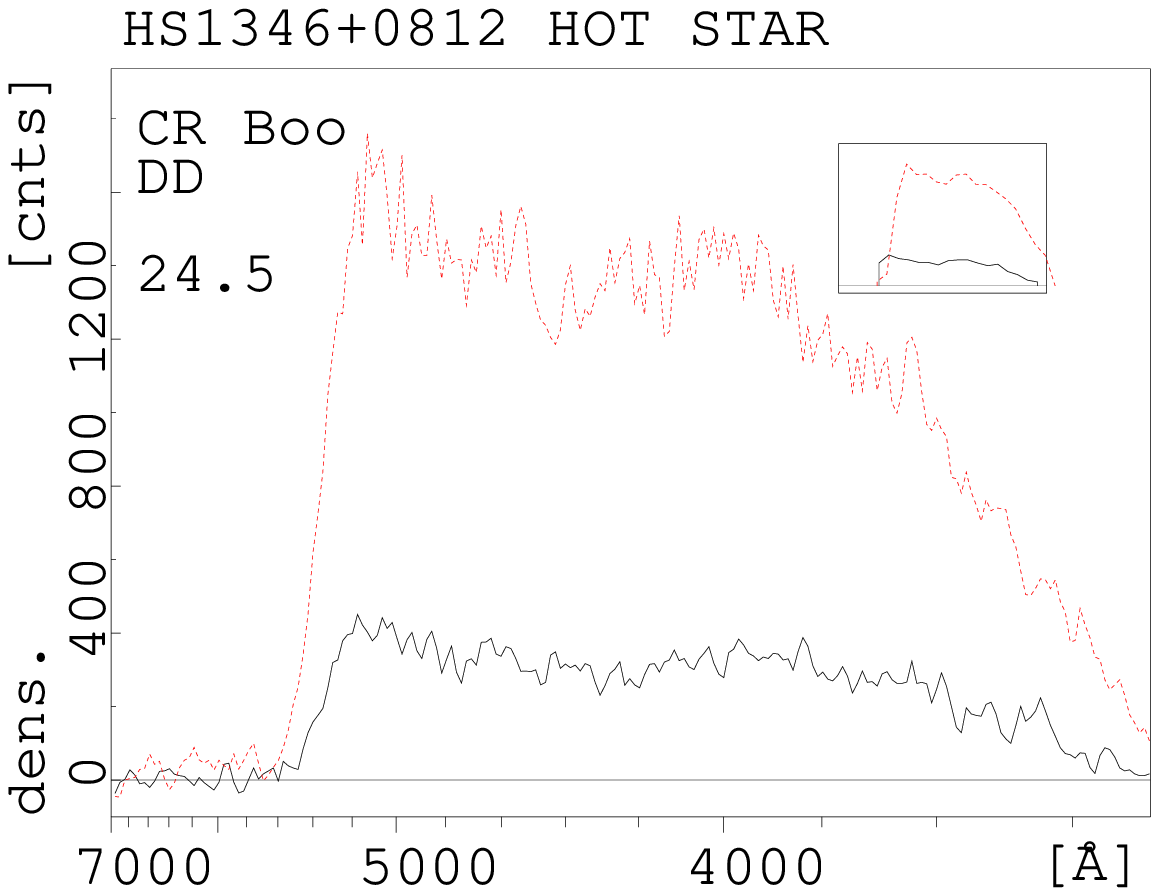,angle=-90,width=4.35cm}}
\mbox{\psfig{file=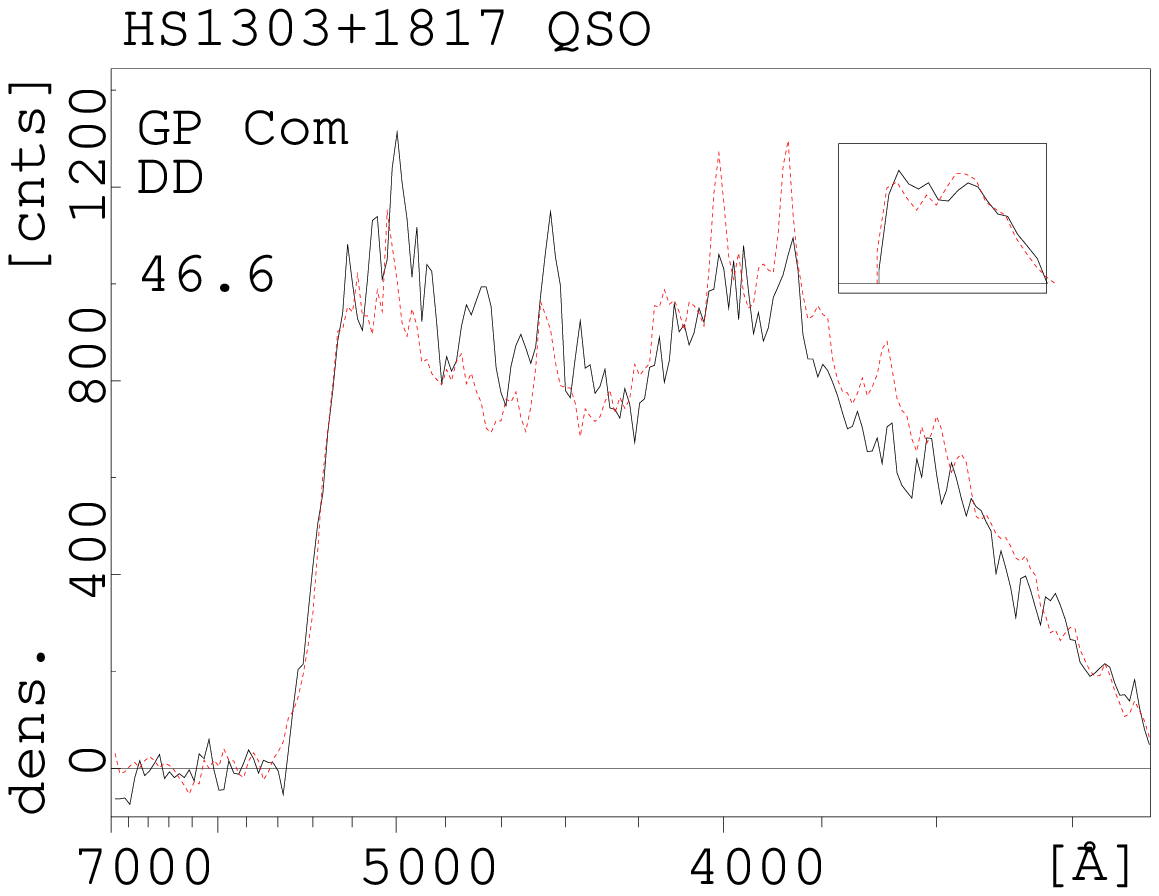,angle=-90,width=4.35cm}}

\caption{HQS HRS density spectra of previously known CVs, orbital
periods are given in minutes. The small inserts show the
low-resolution scans. The classifications based on the visual
inspection of the HRS spectra --~narrow emission line objects, hot
stars, and quasars~-- (Hagen et al. 1999) are given on top of each
panel. For T\,Leo, HT\,Cam, SU\,UMa, PG\,1000+667, CR\,Boo, and
GP\,Com two HRS spectra taken at different epochs are shown.}
\label{f-known_cvs_hrs}
\vspace*{-2ex}
\end{figure}

\section{A large-scale search for new CVs\label{s-cvsearch}}
A handful of CVs were serendipitously discovered in the HQS, the first
ones being the eclipsing dwarf nova EX\,Dra\,=\,HS\,1804+6753
(Billington, Marsh, \& Dhillon 1996) and the possibly magnetic sytem
HS\,0551+7241 (Dobrzycka et al. 1998).
Other examples of rather unusual
CVs found among QSO candidates are two polars with extremely low
accretion rates, HS\,0922+1333 and WX\,LMi\,=\,HS\,1023+3900 
(Reimers, Hagen, \& Hopp 1999; Reimers \& Hagen 2000, see also Schwope
et al., this volume). 11 new CVs were identified using the HQS plate
material for the identification of the optical counterparts of ROSAT
All Sky Survey sources (Bade et al. 1998; Jiang et al. 2000),
including the deeply eclipsing $\Porb=4.2$\,h dwarf nova
GY\,Cnc\,=\,HS\,0907+1902 
(G\"ansicke et al. 2000) and the new SU\,UMa
dwarf nova KV\,Dra\,=\,HS\,1449+6415 
(Nogami et al. 2000).

We have initiated in 2000 a systematic search for new CVs in order to
fully exploit the potential of the HQS for providing a magnitude
limited sample of CVs selected primarily because of their
\textit{spectroscopic} properties.
Figure\,\ref{f-known_cvs_hrs} shows the HQS HRS spectra of 9
previously known CVs, including short and long orbital period dwarf
novae, novalike variables, and double degenerate systems. Balmer
emission, the hallmark of most CVs, can easily be detected in the HRS
spectra for equivalent widths (EW) $\ga10$\,\AA.
At the HRS resolution, novalike variables or dwarf novae in outburst
which are characterised by optically thick disc spectra cannot be
distinguished from hot stars\footnote{Ironically, despite their large
absolute magnitudes, novalike variables are indeed the most
difficult-to-find subtype of CVs, as they are usually neither strongly
variable, nor bright X-ray sources.  The recent discovery of a bright
($V\simeq12.6$!) novalike variable (Mickaelian et al. 2001) underlines
the possible incompleteness of this class of objects.}.  However, at
least two objective-prism plates have been obtained for each HQS field
at different epochs, so that additional information is available on
the variability of the objects.

\begin{figure}[t]
\noindent
\mbox{\psfig{file=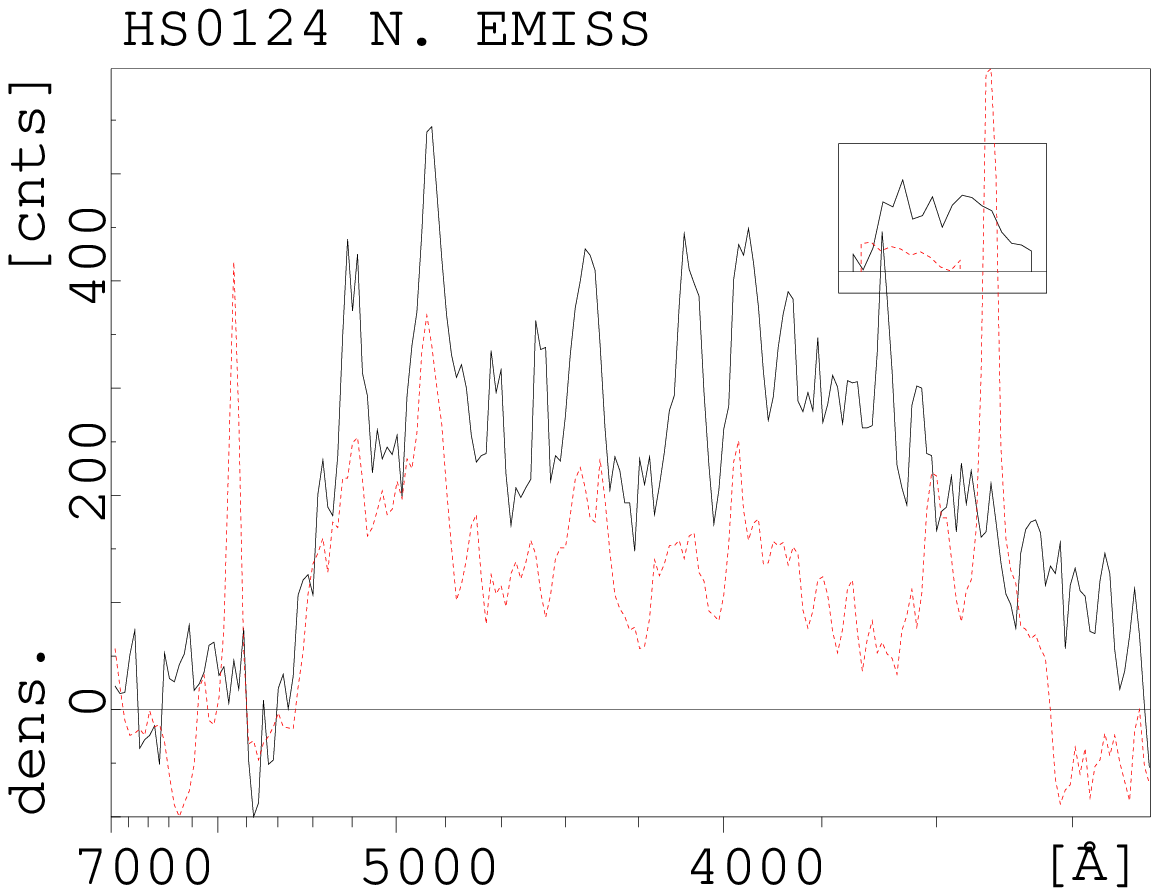,angle=-90,width=4.35cm}}
\mbox{\psfig{file=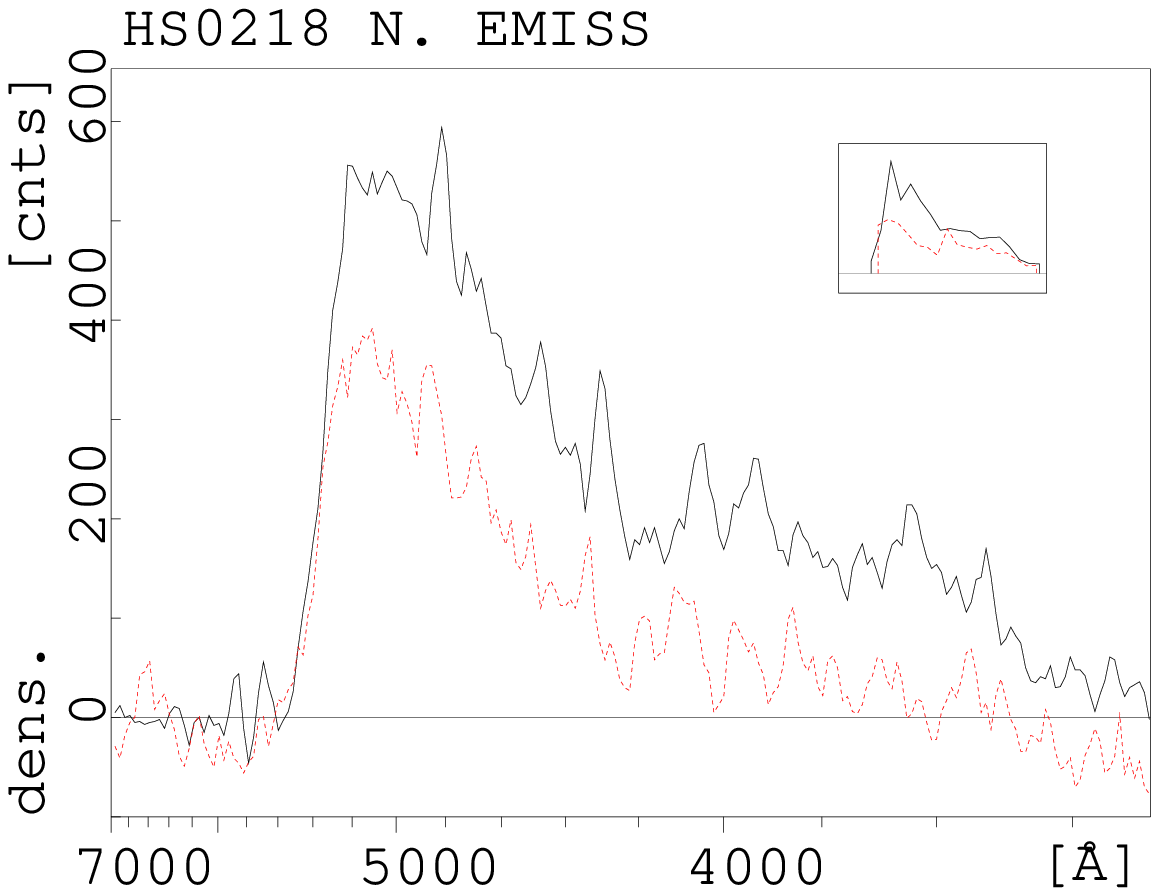,angle=-90,width=4.35cm}}
\mbox{\psfig{file=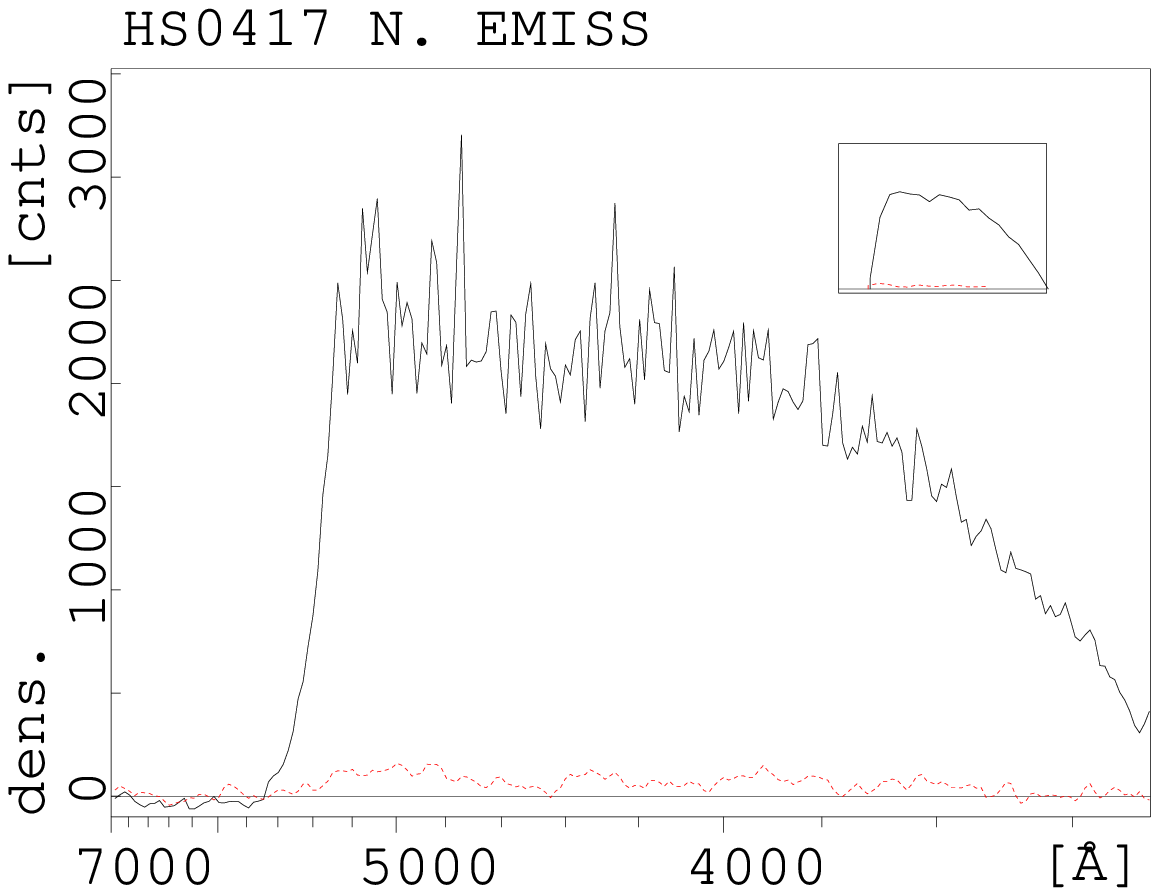,angle=-90,width=4.35cm}}
\caption{HRS spectra of three new CVs discovered in the HQS.}
\label{f-new_cvs_hrs}
\vspace*{-2ex}
\end{figure}

Our criterion to select CV candidates from the HQS is, hence, (1)
detection of Balmer emission\footnote{Note that the same applies of
course for He emission lines in double degenerate CVs, see GP\,Com in
Fig.\,2. A dedicated search for helium cataclysmics is in
preparation.} in the HRS spectrum, or (2) hot star appearance plus
significant variability ($\Delta V\ga1.5$\,mag).  Follow-up
spectroscopy of the CV candidates was obtained during two runs in
September 2000 and April 2001 at the Calar Alto 2.2m telescope using
CAFOS. At the time of writing, 50 CVs have been discovered in the HQS
and 12 very promising CV candidates await spectroscopic confirmation
(e.g., Fig.\,\ref{f-new_cvs_hrs}).

\section{Selection effects in the HQS CV sample}
The characteristics of the HQS (large area, moderately bright cut-off
magnitude) allows a careful examination of the CV detection efficiency
based on the sample of previously known CVs.  The Downes et al. (2001)
catalogue and the TPP database (Kube, G\"ansicke, \& Hoffman, this
volume, p.\,\pageref{paper_kube_tpp}) list 288 CVs or CV candidates
within the area covered by the HQS. This sample can be broken up into
two groups. 

\begin{figure}[t]
\noindent
\mbox{\psfig{file=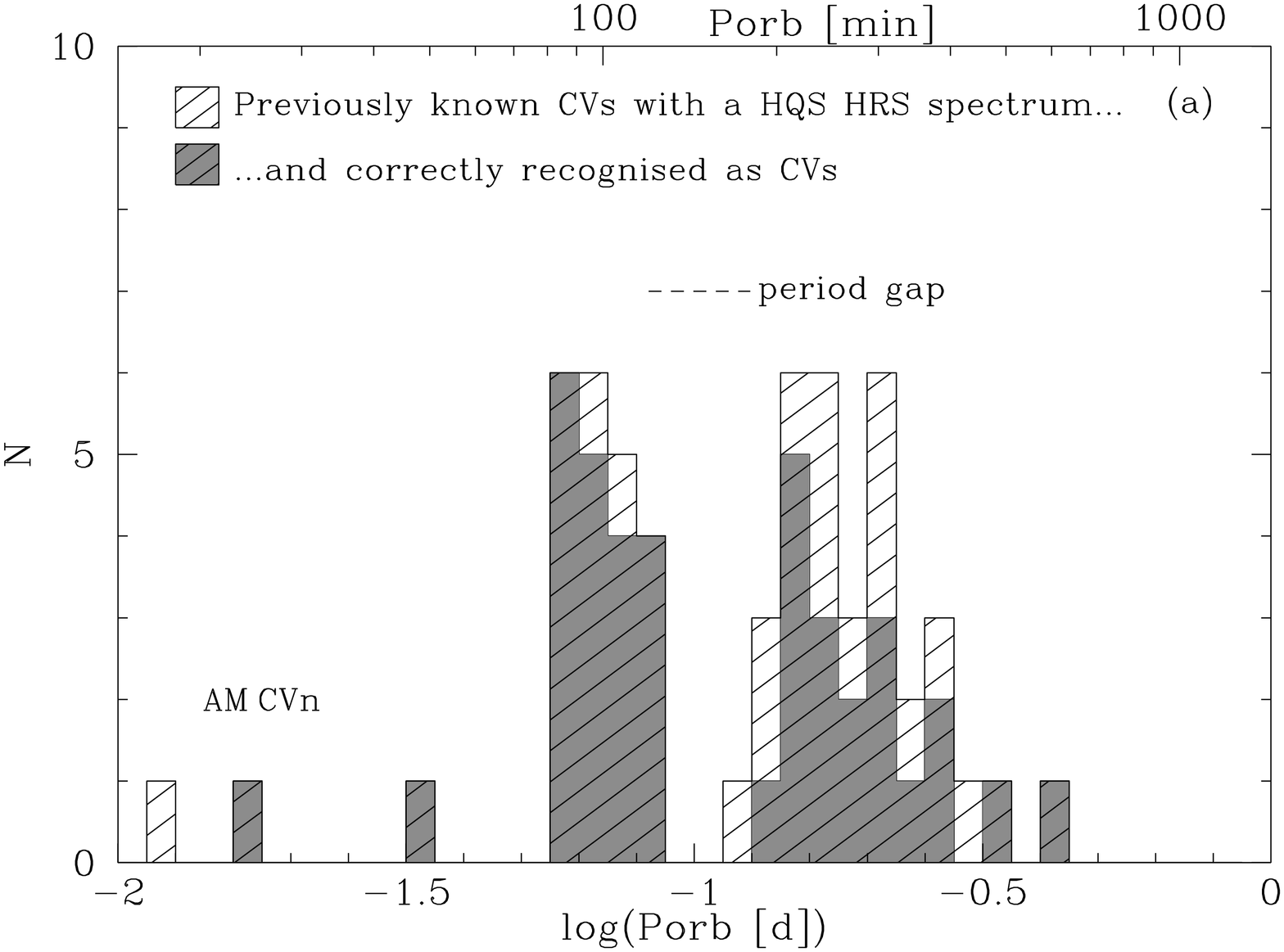,angle=-90,width=6.5cm}}
\mbox{\psfig{file=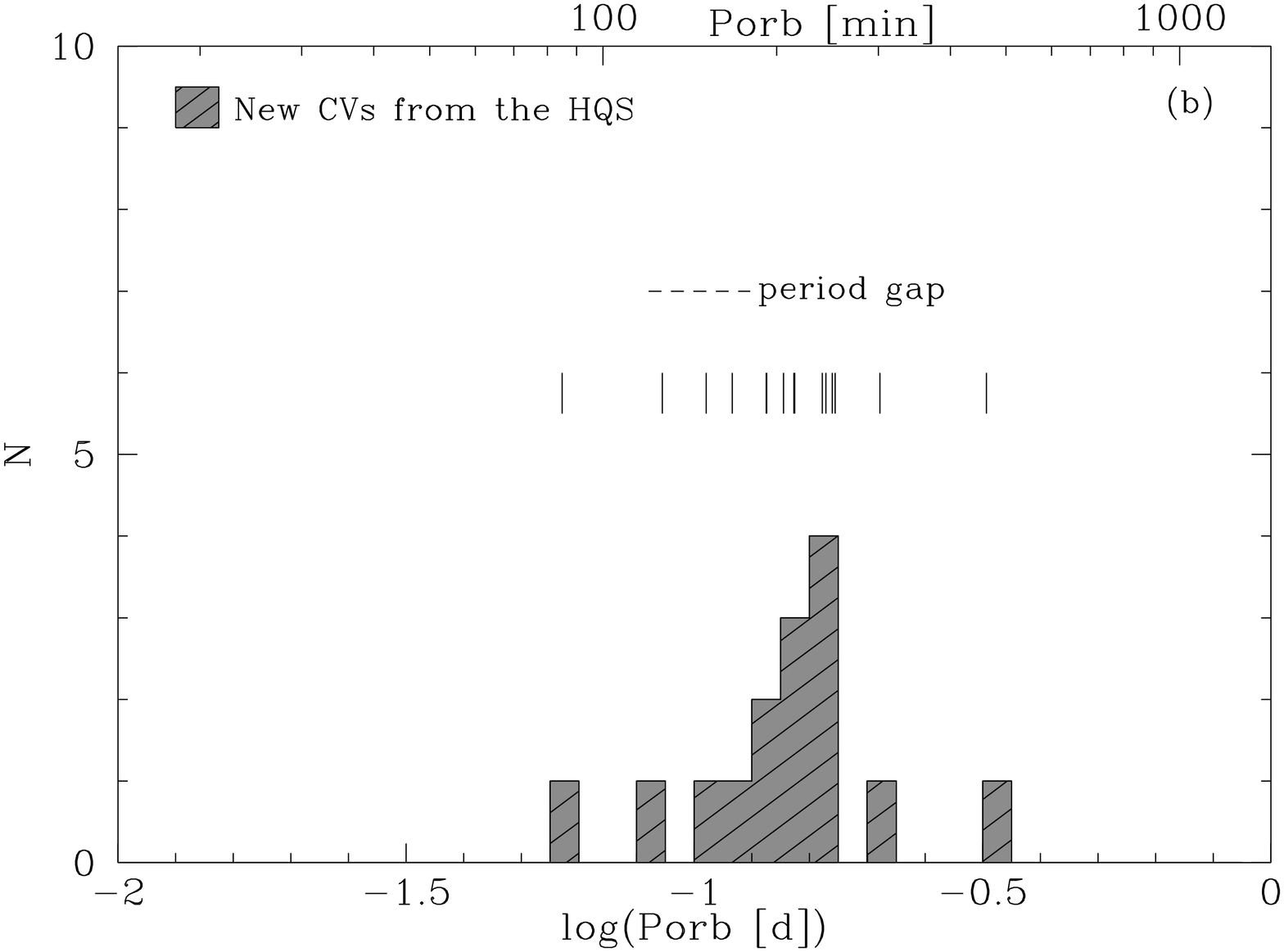,angle=-90,width=6.5cm}}\\

\noindent
\mbox{\psfig{file=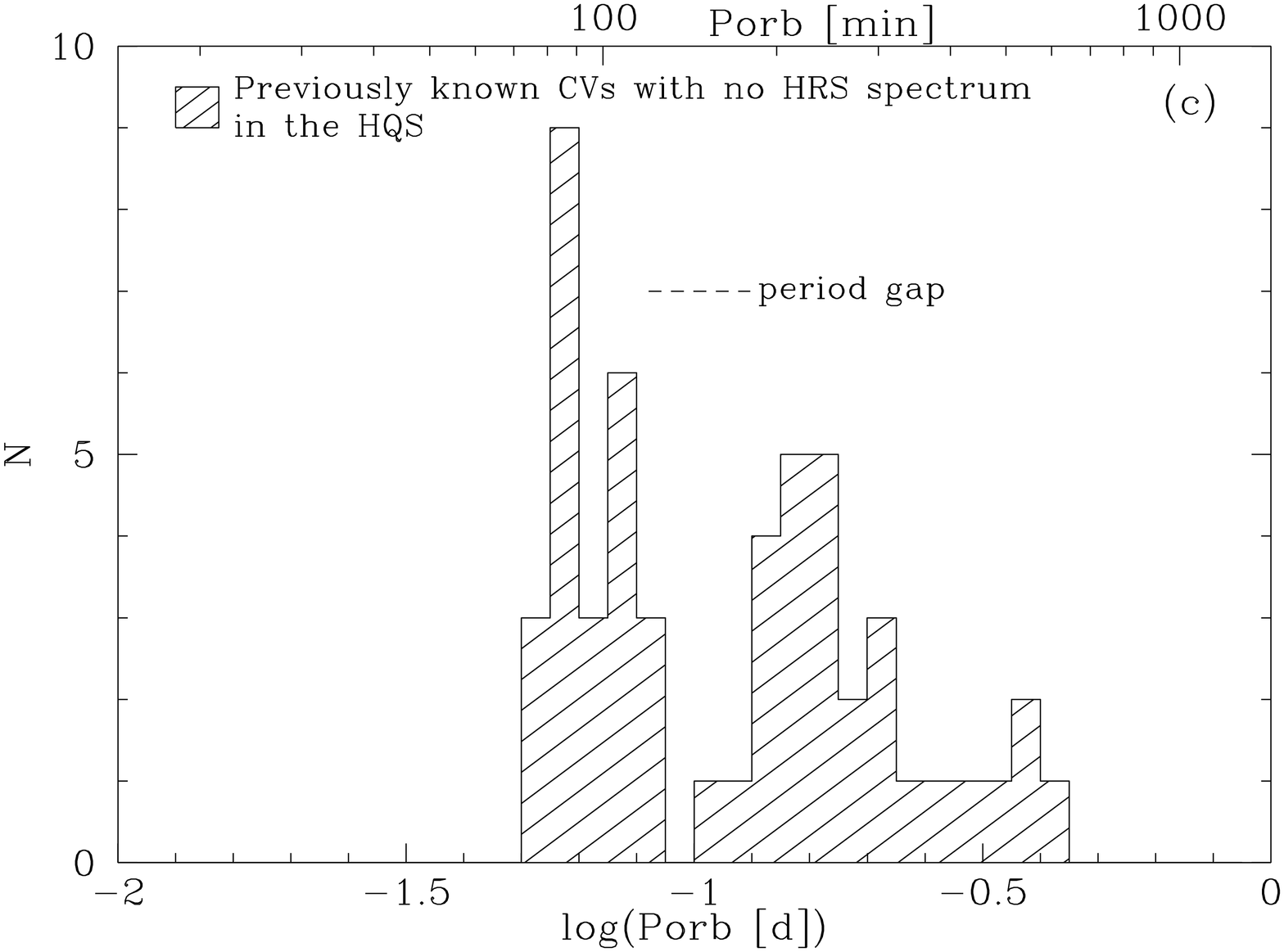,angle=-90,width=6.5cm}}
\hspace*{-0.6cm}
\parbox[b]{7.7cm}{\caption{\label{f-porbs} 
{\bf(a)} Period distribution of the previously known CVs with a HQS
HRS spectrum (shaded) and those correctly recognised as being a CV
(gray). {\bf(b)} Period distribution of the new CVs discovered in the
HQS. The tickmarks indicate the individual periods. {\bf(c)} Period
distribution of the previously known CVs with no HRS spectrum.}
\vspace*{2.5ex}}
\vspace*{-2ex}
\end{figure}

\begin{figure}[t]
\noindent
\mbox{\psfig{file=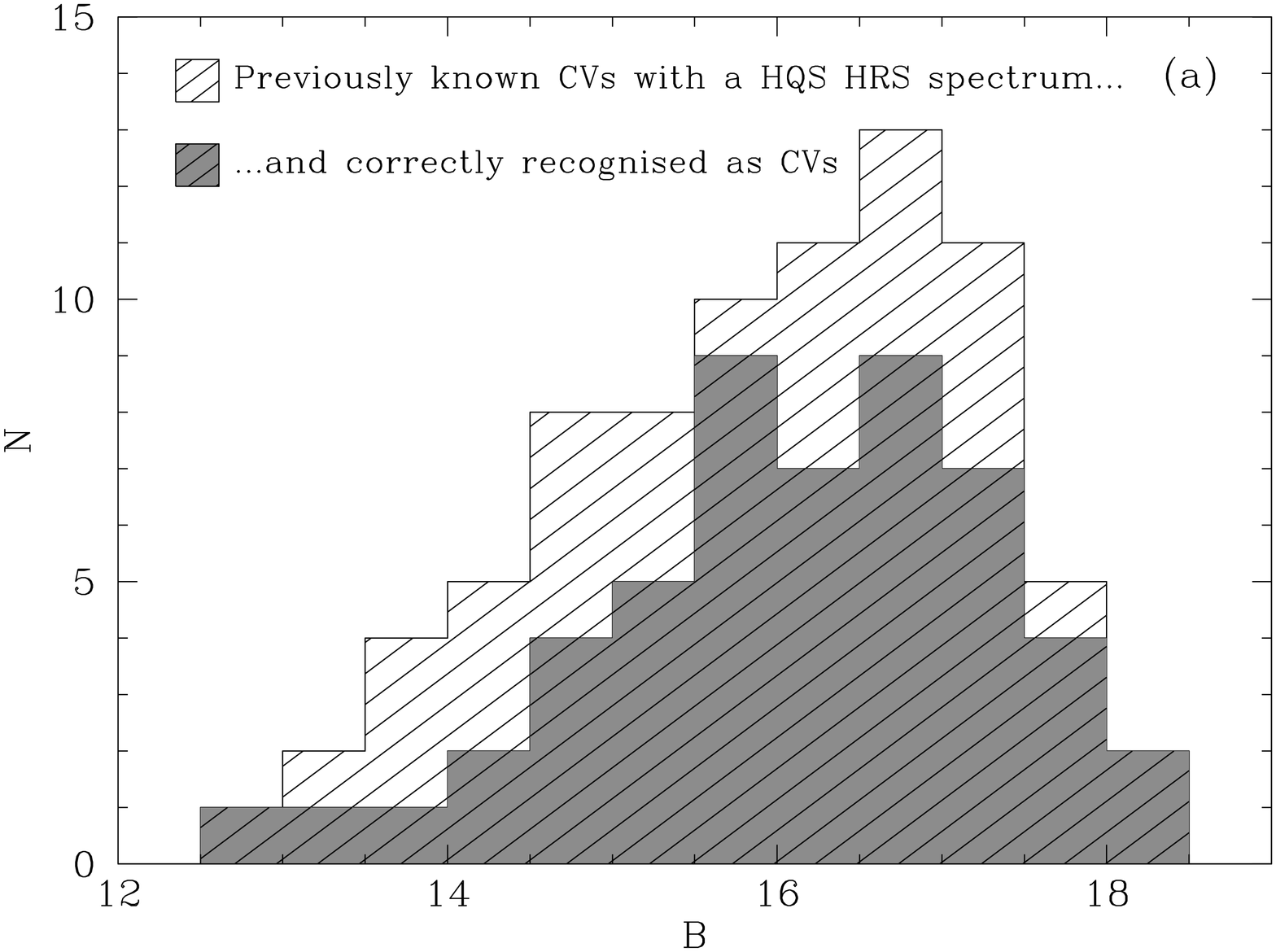,angle=-90,width=6.5cm}}
\mbox{\psfig{file=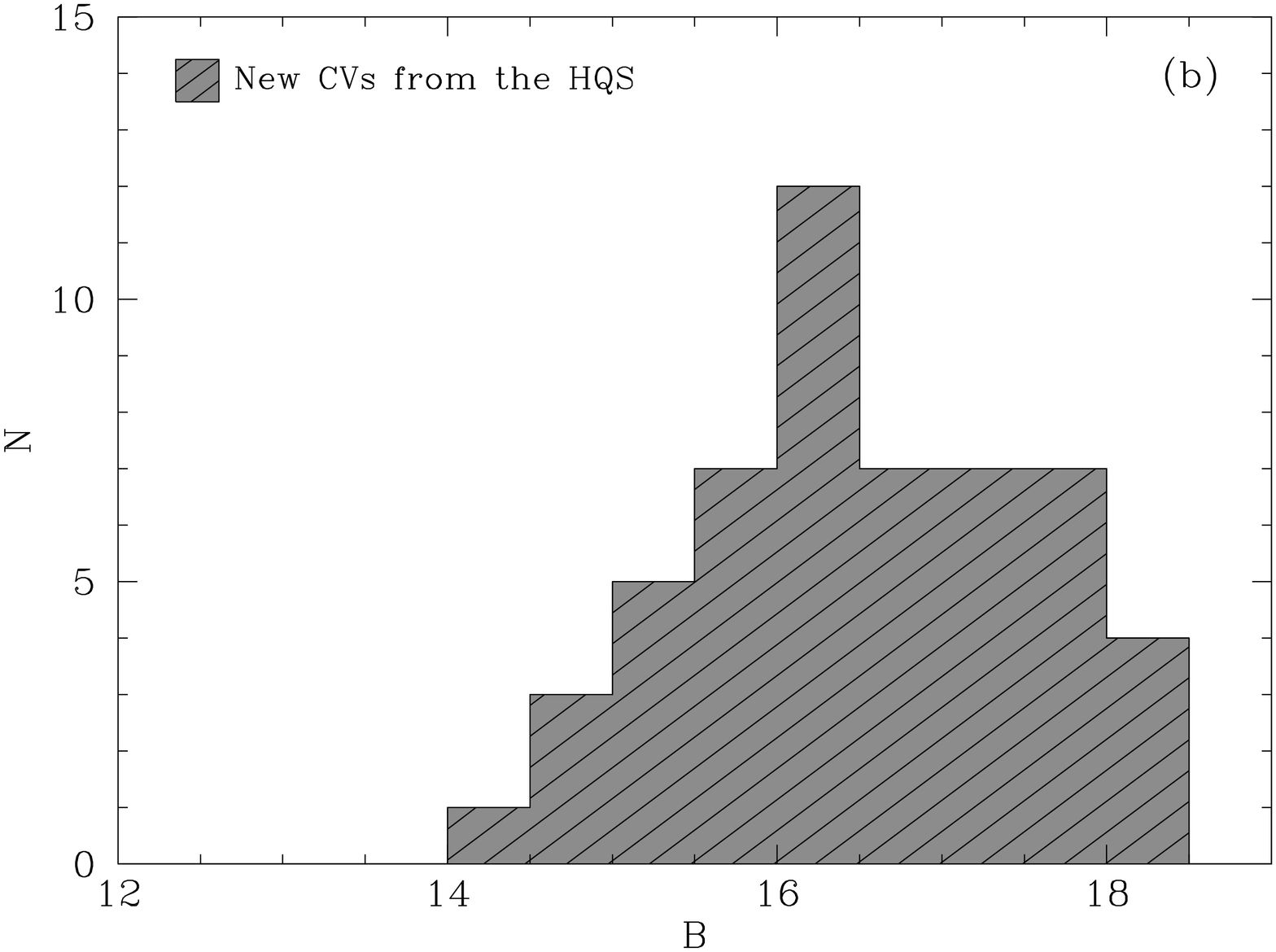,angle=-90,width=6.5cm}}\\

\noindent
\mbox{\psfig{file=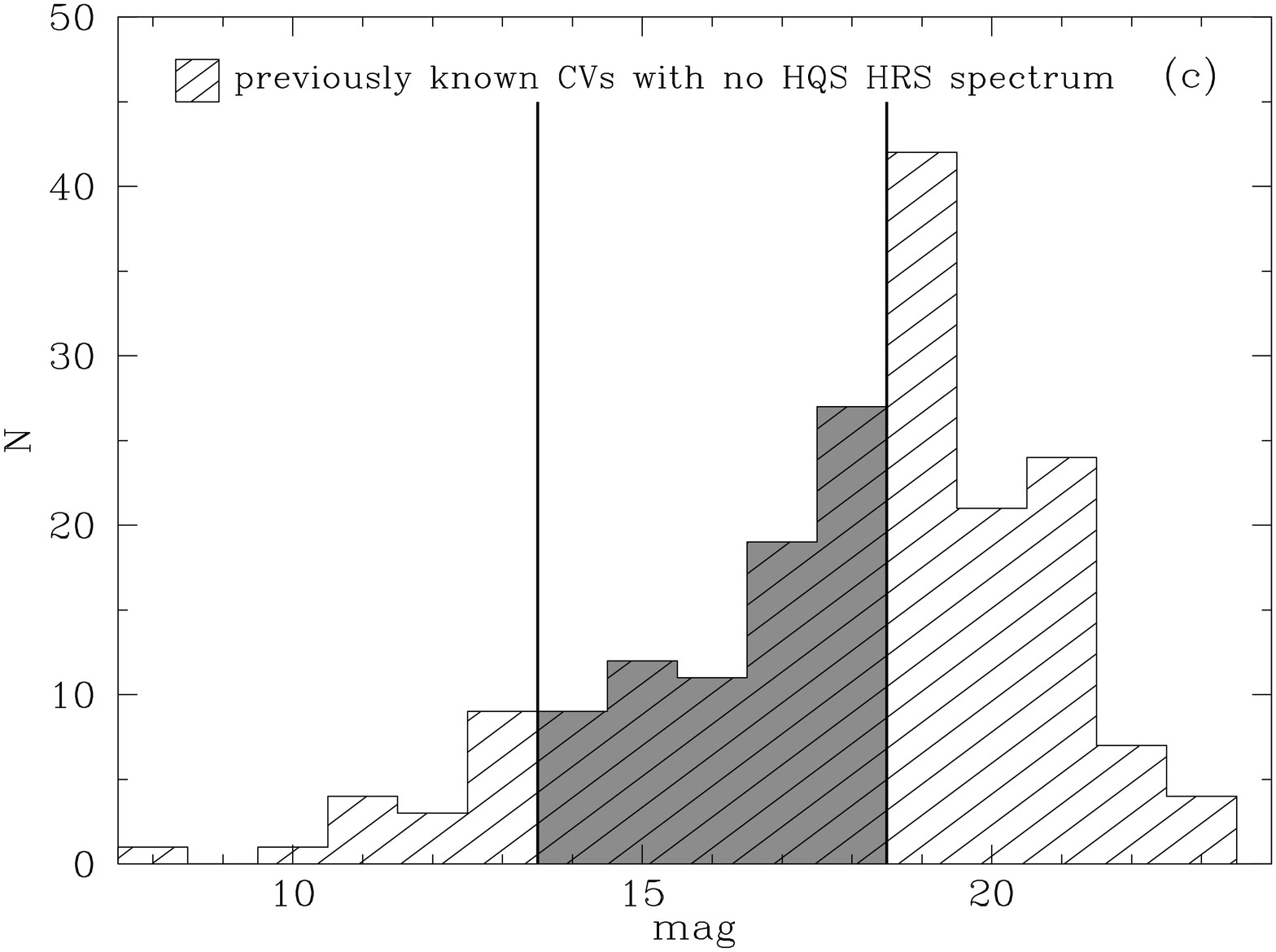,angle=-90,width=6.5cm}}
\hspace*{-0.6cm}
\parbox[b]{7.7cm}{\caption{\label{f-mags}
(a): magnitude distribution of the
known CVs with a HQS HRS spectrum (shaded) and those re-identified as
being a CV (gray). (b): magnitude distribution of the CVs discovered
in the HQS. (c): magnitude distribution of the known CVs
with no HRS spectrum. The dynamic range covered by the HQS is
indicated by the two vertical lines. }
\vspace*{2.5ex}}
\vspace*{-2ex}
\end{figure}

\textit{$\bullet$ Known CVs with a HRS spectrum:}
For 84 out of the 288 CVs in the HQS area at least one density
spectrum is contained in our HRS database. Our selection criterion,
Balmer emission lines or blue continuum plus variability, correctly
recognises 52 of these systems, or 62\,\%, as CV candidates.  The
majority of the CVs that were not recovered are novalike variables
with inconspicuous hot star-like HRS spectra and insignificant
variability information. 
The orbital period has been measured for 57 out of the 84 known CVs,
which allows us to test for a possible selection bias towards short or
long orbital periods. Figure\,\ref{f-porbs} shows the orbital period
distribution of CVs with a HRS spectrum, and, shaded in gray, the
fraction of the CVs correctly recognised by our selection
criterion. Apparently, our detection efficiency is maximal below the
orbital period gap, but we lose a significant fraction of systems
above the gap: the novalike variables with thick disc
emission. Figure\,\ref{f-mags} shows the $B$ magnitude distribution of
the known CVs with a HRS spectrum, and, again shaded in gray, the
fraction of the CVs recovered by our selection. These magnitudes are
derived from the HQS plate material in a homogenous
manner. Interestingly, the fraction of recovered CVs is lowest at the
bright end of the magnitude distribution ($B\sim14$), which is due to
the relatively bright long period novalike variables that could not be
identified as CVs on the base of their HRS spectra.

\textit{$\bullet$ Known CVs with no HRS spectrum:}
A significant fraction (204, corresponding to $\approx60$\,\%) of the
previously known CVs and CV candidates in the HQS area are not
included in the HRS database.  Figure\,\ref{f-mags} shows the
magnitude distribution of these systems, assembled from Downes et
al. (2001) and TPP (Kube et al., this volume). We caution that this
diagram has to be interpreted with some care: the magnitudes are a mix
of $B$, $V$, $pg$, $R$, and $I$ measurements, and, more importantly,
for a significant fraction of the systems (85, corresponding to
$\approx40$\,\%), the published magnitudes are ``fainter than''
limits. With this caveat in mind, Fig.\,\ref{f-mags} suggests that
$\approx20$ CVs are too bright and should be saturated on the HQS
plates, $\approx80$ have magnitudes within the HQS dynamic range, and
$\approx100$ are too faint.

The relatively large number of known CVs and CV candidates with
published magnitudes in the HQS dynamic range that \textit{are not}
included in the HRS database could suggest that the selection
criterion of Hagen et al. (1999), optimized for QSOs, ``missed'' a
significant number of CVs. In order to assess the possibility of a
systematic loss, we carefully inspected the nature of the 80 objects
with no HRS spectrum, as well as the original plate material at the
corresponding coordinates. It turns out that: (a) The majority of the
objects with published magnitudes $\ga17$ are either very faint or not
detected at all on the HQS plates. This is not too surprising, as the
quiescent magnitudes of these poorly studied systems are often
uncertain and the plate limits vary from field to field in the range
$B=18.5\pm0.5$ mag. (b) About a dozen objects with uncertain
identifications/coordinates have no counterpart on the HQS plates,
which suggests wrong coordinates or spurious CV candidates.  (c) A
handful of objects in dense fields were lost due to overlap of the
prism spectra. (d) $\sim10$ CVs were observed during an outburst or a
high state (e.g. ER\,UMa, Z\,Cam, AT\,Cnc), and, hence, their spectra
are saturated on the Schmidt plate. (e) Two objects were scanned in
the high resolution mode, but later on classified as horizontal branch
stars and consequently removed from the HRS database (BK\,Lyn, X\,Leo)

We conclude from our analysis of the known CV/CV candidate sample
that: (1) $\sim90$\,\% of the known CVs with a secure identification
and a quiescent magnitude $13.5\la B \la17.5$ are included in the HRS
database.  For fainter quiescent magnitudes ($17.5\la B\la18.5$), the
fraction of systems with a HRS density spectrum drops rapidly. (2) Our
selection scheme correctly identifies practically all CVs with
emission line $\mbox{EWs}\ga 10$\,\AA\ on the base of their HRS density
spectra. Novalike variables and dwarf novae caught in outburst have
density spectra very similar to hot stars, but may be selected as CV
candidates because of their variability.

\begin{figure}[t]
\noindent
\centerline{\mbox{\psfig{file=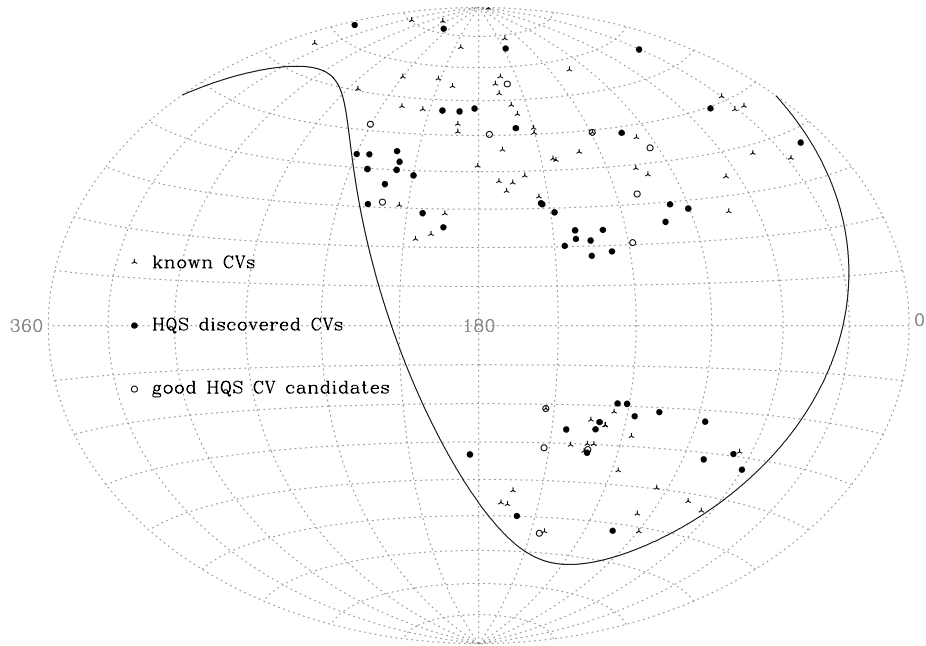,angle=-90,width=12.5cm}}}

\caption{Distribution of the HQS-discovered CVs over the sky (galactic
coordinates, Aitoff projection). The HQS covers the northern
hemisphere ($\delta=0\deg$, indicated by the solid line) and galactic
latitudes $|b|>20\deg$.
\label{f-skydist}}
\vspace*{-2ex}
\end{figure}

\section{Discussion}
At present, 50 CVs have been identified in the course of follow-up
observations of HQS objects, another 12 systems are very good
candidates but need additional spectroscopic confirmation
(Fig.\,\ref{f-skydist}). Compared to the 84 previously known CVs
contained in the HRS database, the new discoveries represent a growth
by $\sim60$\,\% of this sample (location in the HQS area, magnitude in
the range $13.5\la B \la18.5$).
While we are currently refining the process of selecting CV candidates
from the HRS database (including the search for double degenerate CVs
and low mass transfer polars), we are confident that we missed very
few bright CVs ($B\la17.5$) with noticeable Balmer emission lines
(EW$\ga10$\,\AA).  For 15 of the new CVs, orbital period measurements
are available (Fig.\,\ref{f-porbs}). Even though a final discussion
has to await the knowledge of the period distribution of the entire
HQS CV sample, we draw the following --~somewhat preliminary~--
conclusions from the currently available data.

Somewhat surprisingly, only one of the new HQS CVs has an orbital
period below the gap, namely the SU\,UMa type dwarf nova
KV\,Dra\,=\,HS\,1449+6415 ($\Porb=84.93$\,min, Nogami et al. 2000),
with one additional system lying at the lower edge of the gap
($\Porb=126.72$\,min, Thorstensen, private communication). Two more
systems have orbital periods within the gap, and the remaining 11
systems have orbital periods above 3\,h. This result exacerbates the
problem of the ``missing short period CVs'' (Sect.\,1), as our survey
is especially sensitive to the short period systems because of their
characteristic Balmer emission (Sect.\,5, see
Fig.\,\ref{f-porbs}). Considering a conservative faint limit,
$B\la17.5$, our search probes for ``typical'' short period CVs out to
a distance of $\sim150$\,pc, and, so far, our results exclude the
presence of a large population of nearby infrequently outbursting
X-ray faint short period CVs
\textit{unless they have significantly weaker emission lines than,
e.g., WZ\,Sge.}

Deeper surveys (e.g. the 2dF survey [see Marsh this volume,
p.\,\pageref{paper_marsh_2df}] or the Sloan Digital Sky Survey [see
Szkody, this volume, p.\,\pageref{paper_szkody_sdss}]) do find a
larger fraction of short-period systems. However, these high galactic
latitudes surveys with cut-off limits of $V\sim20-21$ reach out to
very large heights $z$ above the Galactic disc, many hundred pc for
long period but low luminosity systems like U\,Gem, and a few kpc for
bright novalike variables like TT\,Ari. Considering that these systems
have a scale height of $\sim150$\,pc (Patterson 1984), it is clear
that the nearby, but intrinsically faint short period CVs overtake the
more luminous long period systems in these deep surveys.

%

While a secure classification of the CVs from the HQS requires
additional follow-up observations, our identification spectroscopy
suggests that we find rather few new polars. This is consistent with
the hypothesis that most bright and nearby polars have been found in
the ROSAT All Sky Survey thanks to their intense soft X-ray emission
(Thomas \& Beuermann 1998).

\vspace*{-2ex}
\section{Conclusions}
\vspace*{-2ex}
We have initiated a large scale search for new CVs, selecting
candidates on the base of their spectroscopic properties in the HQS
HRS database. This selection scheme proved to be very efficient in
recognizing CVs with emission line spectra, while systems with pure
continuum emission can be selected only if sufficient information on
their variability is available. Our search has already significantly
increased the number of moderately bright high galactic latitude CVs
in the northern hemisphere, and produced a number of peculiar CVs that
merit additional detailed observational campaigns, e.g. the eclipsing
systems HS\,0907+1902 (\,=\,GY\,Cnc) and HS\,0455+8315 (G\"ansicke et
al. 2000 and this volume, p.\,\pageref{paper_gaensicke_hs0455}), or
the low mass transfer polars HS\,0922+1333 and HS\,1023+3900 (Reimers,
Hagen, \& Hopp 1999; Reimers \& Hagen 2000; Schwope et al., this
volume, p.\,\pageref{paper_schwope})

The ultimate aim of our project, a detailed confrontation of the
period distribution and the space density of the HQS CV sample with
the predictions of the CV evolution theories and population syntheses
has to be postponed until the system parameters for the entire set of
new CVs have been measured. However, the data collected so far do not
support the high space density of short period CVs predicted by the
standard scenario.

While our survey, along with several other CV searches (see in this
volume: Lott et al., p.\pageref{paper_lott}; Marsh et
al. p.\pageref{paper_marsh_2df}; Szkody et
al. p.\pageref{paper_szkody_sdss}; Tappert et
al. p.\pageref{paper_tappert}), will significantly improve our
knowledge on the actual space densities of CVs, a number of hypotheses
that can avoid too numerous short period CVs have already been
proposed (e.g. Shara et al. 1986; Patterson 1998; King \& Schenker and
Schenker \& King, this volume, p.\,\pageref{paper_king},
\pageref{paper_schenker}). After a long period of an almost dogmatic
standstill, the topic of CV evolution has become a very lively field
again.

\acknowledgements{We would like to thank all the people who
contributed observations to this project: H. Barwig.  R. Fried,
M. Kuduz, D. Nogami, R. Schwarz, A. Staude, P. Szkody.
J. Thorstensen is acknowledged for sharing his orbital periods for a
number of HQS CVs prior to publication.
BTG was supported by DLR/BMBF grant 50\,OR\,9903\,6. The HQS was
supported by the Deutsche Forschungsgemeinschaft through grants Re
353/11 and Re 353/22.}


\vspace*{-2ex}
\begin{references}
\vspace*{-1ex}
\reference Bade, N., Engels, D., Voges, W., et~al. 1998, A\&AS 127, 145
\reference Beuermann, K., Thomas, H.C., Reinsch, K., et~al. 1999, A\&A 347, 47
\reference Billington, I., Marsh, T.R., \& Dhillon, V.S. 1996, MNRAS 278, 673
\reference de Kool, M. 1992, A\&A 261, 188
\reference Dobrzycka, D., Dobrzycki, A., Engels, D., \& Hagen, H.-J. 1998, AJ 115, 1634
\reference Downes, R.A. 1986, ApJ 307, 170
\reference Downes, R.A., Webbink, R.F., Shara, M.M., et~al. 2001, PASP 113, 764
\reference G\"ansicke, B.T., Fried, R.E., Hagen, H.J., et~al. 2000, A\&A 356, L79 
\reference Green, R.F., Ferguson, D.H., Liebert, J., \& Schmidt, M. 1982, PASP 94, 560
\reference Hagen, H.J., Groote, D., Engels, D., \& Reimers, D. 1995, A\&AS 111 195 
\reference Hagen, H.J., Engels, D., \& Reimers, D. 1999, A\&AS 134, 483
\reference Hertz, P., Bailyn, C.D., Grindlay, J.E., et~al. 1990, ApJ 364, 251
\reference Howell, S.B., Rappaport, S., \& Politano, M. 1997, MNRAS 287, 929
\reference Jiang, X.J., Engels, D., Wei, J.Y., Tesch, F., \& Hu,
   J.Y. 2000, A\&A 362, 263
\reference Jordan, S. 1997, in White Dwarfs, eds. J. Isern, M. Hernanz,
  \& E. Garc{\'\i}a-Berro (Dordrecht: Kluwer), 397
\reference King, A.R. 1988, QJRAS 29, 1
\reference Kolb, U. 1993, A\&A 271, 149
\reference  Mickaelian, A.M.,   Balayan, S.K.,  Ilovaisky, S.A., et
 al. 2001, A\&A in press
\reference Nogami, D., Engels, D., G\"ansicke, B.T., et~al. 2000, A\&A 364, 701
\reference Patterson, J. 1984, ApJS 54, 443
\reference Patterson, J. 1998, PASP 110, 1132
\reference Politano, M. 1996, ApJ 465, 338
\reference Reimers, D. \& Hagen, H.J. 2000, A\&A 358, L45
\reference Reimers, D., Hagen, H.J., \& Hopp, U. 1999, A\&A 343, 157
\reference Ritter, H. 1986, A\&A 168, 105
\reference Ringwald, F.A. 1996, in Cataclysmic Variables and Related
 Objects, eds. A. Evans \& J.H. Wood (Dordrecht: Kluwer), 89
\reference Shara, M.M., Livio, M., Moffat, A.F.J., \& Orio, M. 1986, ApJ 311, 163
\reference Thomas, H.C., Beuermann, K., Reinsch, K., et~al. 1998, A\&A 335, 467
\reference Thomas, H.C. \& Beuermann, K. 1998, Lecture Notes in
   Physics 506, 247
\reference Warner, B. 1974, MNAS 33, 21
\reference Wickramasinghe, D.T. \& Ferrario, L., 2000, PASP 112, 873
\end{references}
\end{document}